\documentclass[preprint, times, a4paper,onecolumn,oneside,12pt,review]{elsarticle}
\usepackage{IEEEtrantools}
\usepackage{algorithm, algpseudocode,graphicx}
\usepackage{amsmath,amsthm,amssymb,amsfonts,latexsym,mathrsfs}

\usepackage{setspace}

\theoremstyle{definition}
\newtheorem{theorem}{Theorem}

\newtheorem{proposition}[theorem]{Proposition}
\newtheorem{example}{Example}
\def\calh{\mathcal{H}}
\def\calu{\mathcal{U}}
\def\calf{\mathcal{F}}
\def\boldy{\mathbf{y}}
\def\myactive{\mathit{active}}
\def\prob{MAR-constrained hole assignment problem}
\def\probcap{MAR-Constrained Hole Assignment Problem}
\def\oracle{\textsc{SEP2-DP-Oracle}}
\def\bfreturn{\textbf{return}}
\def\bfsr{\textbf{store and return}}

\def\bfand{\textbf{and}}
\def\acro{MC-HAP}
\def\len{\mathrm{len}}
\def\bfi{\mathbf{I}}
\def\bfy{\mathbf{y}}
\def\bfz{\mathbf{z}}
\def\bnp{B\&P}
\def\bnb{B\&B}
\def\sepi{\textsf{B\&P-SEP1}}
\def\sepii{\textsf{B\&P-SEP2}}
\def\their{\textsf{LAM-T2}}

\usepackage{xcolor}
\usepackage{hyperref}

\usepackage[margin=1in]{geometry}
\usepackage{xcolor}
\date{}
\journal{Ad Hoc Networks}

\usepackage{fancyhdr}
\begin{document}
\lhead{\footnotesize Accepted Manuscript - Ad Hoc Networks - \url{https://doi.org/10.1016/j.adhoc.2022.102871}}
\rhead{\thepage}
\cfoot{\color{blue}\textcopyright{} 2022. This manuscript version is made available under the CC-BY-NC-ND 4.0 license \url{https://creativecommons.org/licenses/by-nc-nd/4.0/}}

\begin{frontmatter}
\title{A Branch-and-Price Approach to 
	a Variant of
	the
	Cognitive Radio Resource Allocation Problem\tnoteref{t1}\tnoteref{t2}}
\tnotetext[t1]{This research is funded by Iran's National Elites Foundation (INEF).}
\tnotetext[t2]{Declarations of interest: none.}

\author[1]{Hossein~Falsafain\corref{cor1}}
\ead{h.falsafain@iut.ac.ir}
\author[1]{Mohammad~Reza~Heidarpour}
\ead{mrheidar@iut.ac.ir}
\author[2]{Soroush Vahidi}
\ead{sv96@njit.edu}

\cortext[cor1]{Corresponding author}
\address[1]{Department of Electrical and Computer Engineering, Isfahan University of Technology, Isfahan 84156-83111, Iran}
\address[2]{Department of Computer Science, New Jersey Institute of Technology, Newark, NJ, USA}

	\begin{abstract} \small
Radio-frequency  portion  of
the electromagnetic spectrum is 
a scarce resource.
Cognitive radio (CR) technology
has emerged as
a promising solution
to 
overcome
the spectrum
scarcity bottleneck.
  Through this
technology,
secondary users
(SUs)
sense the 
spectrum opportunities free from
primary users (PUs),
and opportunistically 
take advantage of these (temporarily) idle portions, known as spectrum holes. 
In this correspondence,
we consider a variant of
the cognitive
radio 
resource allocation problem
posed by 
Martinovic~\textit{et~al.}~in
2017.
 The distinguishing feature of this
 version of the problem is that
 each SU, due to its hardware limitations,
 imposes  the requirement that
 the to-be-aggregated 
 spectrum holes
 cannot be arbitrarily far from each other.
 We call this 
 restriction
 as
 the Maximal
 Aggregation Range (MAR) constraint, and refer to this variant of the problem as 
 the \prob{}.
 The
  problem
 can be formalized as an
 NP-hard
 combinatorial optimization problem.
		 We propose a novel binary
		integer linear programming (ILP)
		 formulation to the problem.
		The number of constraints in this
		 formulation is
		the number of spectrum holes plus
		the number of SUs.
		On the other hand,
		the number of binary decision variables
		in the formulation can
		be prohibitively large,
		as for each legitimate spectrum allocation to each SU, one variable is needed.
		Due to this difficulty, the
		resulting
		 0-1~ILP
		instances
		may not even be
		explicitly describable,
		let alone be solvable by an ILP solver.
		We propose a 
		branch-and-price
		(\bnp{})
		 framework to tackle this
		challenge.
		This framework is in fact a
		branch-and-bound procedure in which at each
		node of the search tree, we utilize 
		the so-called (delayed) column generation
		technique for solving
		the LP relaxation of the corresponding subproblem.
		%
		%
%For solving
%		the pricing problem,
%				we have used the
%		dynamic programming approach.
		%
		% 
As evidenced by the numerical results,
the LP relaxation bounds are
very
 tight.
%The absolute integrality gaps
%are in general very small.
 This allows for  a very effective
pruning of the search space.
Compared to the previously suggested formulations,
the proposed technique
can
 require much less computational effort.
%
%
% In particular, our simulation results show that for
%problem instances consisting of
%?? spectrum holes and
%?? SUs,
% the proposed \bnp{} approach
%  needs ?? times less
%  CPU time compared to
%   the best presently available ILP formulation.
	\end{abstract}
%	%
%	

\end{frontmatter}
\thispagestyle{fancy}

	\section{Introduction} 	
  Radio-frequency  portion  of
the electromagnetic spectrum,
which ranges
from
about 3~kHz to 300~GHz,
is inevitably a limited resource,
and a
huge amount of
it 
considered as exploitable
has already been
allocated to licensed owners.
 Today's ever-increasing
 demand for radio
 resources,
 which is
 likely to outstrip the available spectrum, calls for novel
 ideas and techniques to overcome the traditional barriers in spectrum exploitation. 
  One notable fact is that, most of the licensed radio spectrum is severely
 underutilized \cite{Mishra04,Chiang2007,fcc,ofcom,Haykin05}.
 As reported by the Federal Communications Commission (FCC), the shortage of spectrum
 resources can mainly
 be attributed to the  
 \textit{spectrum access
 	problem}, rather than the \textit{spectrum crowdedness} \cite{fcc}. 
In response to the
spectrum scarcity
challenge, the idea of using cognitive radios
(CRs), which
was first proposed by Mitola and Maguire in
1999~\cite{Mitola99},
has been found to be a promising approach.
 In the hierarchical access model for dynamic spectrum access,
 the spectrum users are classified into
 two types: licensees and 
 non-licensees~\cite{survey.dsa}.
   A licensee 
 has the exclusive right to access its
 licensed band. 
 In the jargon of CR, it is commonly referred to as a \textit{primary user} (PU),
 and accordingly, a non-licensee
 is called a  \textit{secondary user} (SU). 
% A PU is also known as
% a licensed, incumbent,
% or non-cognitive user,
% and an SU is also known as
% a cognitive or unlicensed user.
 CR is, in fact, a paradigm
 for enabling the spectral coexistence
 between the PUs and the SUs.
 The PUs have to be
 protected from harmful interference,
 i.e., the activities of the SUs
 have to be made transparent
 to them.

    Over the time, different approaches
 have been suggested
 to establish coexistence between
 PUs and SUs.
 In all of these approaches,
 the ultimate goal is to provide
 spectrum access to
 SUs without
 introducing any disturbance to the
 normal operation of PUs.
 In the 
 %first and 
 most
 widely adopted approach,
 known as the
 \textit{interweave} paradigm,
 the aim is to
 provide SUs with an
 \textit{opportunistic} spectrum
 access.
 Therefore, SUs
 utilize the temporarily
 vacant portions of 
 the spectrum, referred to as  \textit{spectrum holes} (or white spaces). 
   CR technology has its own intricacies and challenges. 
   Efforts
   are needed to address these
   challenging problems.
 One of these challenges is
 to perform spectrum allocation to
 SUs in an effective manner.
 {Several variants of 
 the problem of spectrum allocation,
 also referred to as spectrum
 assignment
  and frequency assignment,
 have already been
 investigated in the
 literature
 (see \cite{Tragos13,Hu2018,survey3} and references therein).
 In this contribution,
 we consider a variant of
 the problem first
 introduced  by 
 Martinovic~\textit{et~al.}~in \cite{ilp}.
 The distinguishing feature of this
 version of the problem is that, for each
 SU,
 it takes into account the 
 requirement of
 closeness of the
 to-be-aggregated spectrum holes.}
 In fact,
 since a single 
 spectrum hole may be too small to
 satisfy the spectrum demand of an
 SU,
 we have to resort to technologies
 such as software defined radio 
 to glue these unoccupied intervals
 together.
 However, only a few
 research works
 have taken users' hardware limitations
 into account when aggregating spectrum holes~\cite{ilp}. 
 In fact, it may be the case that 
 the spectrum holes are spread across a wide frequency range,
 but due to
 certain technological restrictions,
 only those
 that lie in a certain
 specified distance of each other
 can be considered to be aggregated. 
 This practical concern can be taken into account by considering per-user 
 upper-bound constraints in the
 problem formulation. 
 As in~\cite{ilp},
 we will
  refer to
 this per-user upper-bound
  as the 
 \textit{maximal aggregation range~(MAR)}.
 The MAR constraints
 were taken into account 
 for the first time in~\cite{Martinovic16},
 but for the special case of SUs
 of the same spectrum requirement.
  In~\cite{ilp},
 the authors have considered
 a more general setting in which
 the SUs are heterogeneous
 in the sense that 
 they 
 may have different
 spectrum demands and MARs.
 They have proposed
 two different 0-1 
 integer linear
 programming
 (ILP) formulations
 for this more general version of the
 problem.  
 The resulting models can then
 be fed to a 
 general-purpose ILP solver (e.g., CPLEX and GUROBI).

Typically, CR resource management
problems are computationally hard.
This is also the case for the problem
considered in this paper.
 %%%
 Methods for solving
 computationally hard
 problems can
 generally
  be categorized 
  into two main groups:
   exact and inexact. 
   An exact approach
   is  guaranteed to return an optimal
 solution if  
 one exists (see, e.g., 
 \cite{ilp,Martinovic16}). 
 An inexact approach, on the other hand,
  can return a 
 satisfactory%
  ---hopefully optimal or near-to-optimal---%
  solution
  in a reasonable (polynomial) time
  (see, e.g., \cite{mathprog-2,mathprog-5}).
 %
 %
%   For instance,
% branch-and-bound~(\bnb)~\cite{exact-1}
% is an exact,
% and genetic algorithm~\cite{???}
%  is an inexact
% method
% for solving NP-hard optimization problems.
 The branch-and-price~(\bnp{}) routine described
 in the present  paper is
 an exact approach.

 Tools and
 techniques
 originating
 from the
 discipline of
 mathematical 
 programming
 have proven
 to be
 very useful
 in
 solving
 computationally hard 
 CR
  resource management
  problems,
  in both exact and inexact manners~(see, e.g.,~\cite{ilp,Martinovic16,mathprog-2,mathprog-5}).
Specifically,
  many computationally hard 
  combinatorial (discrete)
  optimization problems
    are
  naturally expressible as
   integer linear
  programs~\cite{ilp,Martinovic16}. 
 Generally, such a problem,
 can be reduced (formulated, modeled)
 as an integer linear
 program
 in various ways~\cite{ilp,Martinovic16,applied}.
 The main
 accomplishment of this paper
 is, firstly, the introduction of a novel 0-1~ILP formulation for the
 above-mentioned variant of
 the cognitive radio resource allocation
 problem, which is hereinafter 
 referred to as \textit{ the
  \probcap{} (\acro{})}. 
 The aim is
  to maximize the spectrum utilization
  subject to 
  the constraints imposed by hardware limitations.
 Because of the (potentially) huge number
 of decision variables,
 the 
 associated
 0-1 integer linear programs 
 cannot be described explicitly,
 and therefore
 cannot be fed to an ILP solver.
 Hence,  
 for solving these programs,
 we resort to
 the well-known framework of
 \bnp{}~\cite{dual-2,branching,dual-4,applied}.
 This is, in fact,
 a linear-programming-relaxation-based branch-and-bound
 (\bnb{})
 framework within which
 the linear programming (LP)
 instances are solved using
 the so-called (delayed)
 column generation
  method.
 Our numerical results show that 
 the formulation yields a 
 very  tight (strong)
 LP relaxation
 (but, as stated earlier,
 at the expense of
 a potentially huge number of binary
 decision variables).
 %
 %
% In fact,
%for ???\% of the considered instances,
% the absolute
% integrality gap
%is zero,
% which means that,
% for these 
% 0-1~ILP instances,
% the optimal objective value
% of the LP relaxation 
% (i.e.,
% the value 
% returned by the column generation
% procedure at the root node)
% is equal to the
% optimal objective value
% of the instance itself.
 This leads to a very effective
 pruning of the search space.
% Moreover,
% for
% ???\% of the considered instances,
% the solution to the LP
% relaxation is itself integral, i.e.,
% at the root node,
% the column generation procedure alone
% solves the instance, and
% no branching occurs at all.
 As evidenced by the 
 simulation results,
 the 
 proposed \bnp{} approach
  exhibits a superior performance
  in terms of the
  required CPU time
   compared
 to the best currently available
 0-1~ILP formulation of the problem,
 which is presented
 in~\cite{ilp}.

It is worth noting that 
the \acro{}
has a lot in common
with the Generalized Assignment Problem (GAP). The GAP
can be described as 
finding a maximum profit assignment of 
$ n $ jobs to $ m $ ($ n\geq m $)
 agents such that 
each job is assigned to exactly one agent, 
and that each agent is permitted to be assigned to more than one job, subject to its capacity limitation~\cite{applied,gap-survay,gap-bnp}. 
From the perspective
of the GAP,
each SU can be
seen as an agent,
and each hole can be
seen as a job.
The distinguishing point between
the two problems is 
that in the \acro{},
each SU has its own
MAR,
but in the GAP,
the an  agent
does not impose such a
restriction.
In fact,
in the GAP,
assigning a job
to an agent 
does not prohibit 
the assignment of
another job,
as long as enough capacity
is available.
On the other hand, in 
the \acro{},
the assignment of 
a hole $ h $ to
an SU, forbids us
from assigning
the holes whose
distances
from $ h $
are greater than
what MAR dictates.
Therefore,
the existing 
approaches for solving the
GAP, in
an exact or inexact manner,
although very inspiring,
are not directly
applicable
for the \acro{}.
Moreover,
in the \acro{},
 there is a kind 
of \textit{conflict} between the holes.
Holes that are far apart are in conflict with each other. Therefore, the approaches
available to solve problems
  such as the \textit{bin packing with conflicts problem},
  can be inspiring 
  in the design of algorithms for
  the \acro{}.
  The \bnp{}
  approach has been employed for both of the 
  above-mentioned problems \cite{gap-bnp,bpack-bnp2,bpack-bnp1}.

 The organization of this
 contribution  is as follows. In Section~2, we
 introduce
 some preliminaries and notation,
 and provide a rigorous formulation
 of the \acro{}. Section~3
 is devoted to
 the presentation of
 our novel 0-1~ILP formulation 
 of the problem, along with
 a brief overview of the
 \bnp{} framework.
 In Section~4,
 we describe
 the column generation procedure,
 discuss the
 pricing problems,
 an describe their
 corresponding
 pricing oracles.
 Section~5
 is dedicated to
 experimental evaluations, and to numerical
 comparisons with the best currently available ILP formulation
 of the problem. We extensively compared our \bnp{}
 approach with a
 formulation presented in~\cite{ilp}. Finally, some concluding remarks
 are offered  in Section~6.

 \section{Notations and Problem Statement}	
 
 In this section,
 we provide a formal definition of
 the \acro{}, and  
 introduce some definitions and notations.
 A summary of notations used in this paper
  is shown in Table~\ref{tab.notat}.
 In an instance 
 of the \acro{}, we are given two sets
 $\calh$ and
 $\calu$, where
 $\calh=\{h_1,h_2,\ldots,h_M\}$ is the set of all available
 spectrum holes, and 
 $\calu=\{u_1,u_2,\ldots,u_N\}$
 is the set of all SUs.
 Each spectrum hole $h_i$, 
 $1\leq i \leq M$,
 is specified by its
 left endpoint $\alpha_i$ 
 and right endpoint $\beta_i$,
 i.e., the hole can be represented
 by the interval $[\alpha_i,\beta_i]$.
 We assume that spectrum holes are
 pairwise
  disjoint, and 
  appear in $ \cal H $ based on
 their left endpoints,
 i.e., we have
  $ \alpha_1 < \alpha_2 < \cdots < \alpha_M  $.
   We denote the length of the hole
  $h_i$, i.e., $\beta_i-\alpha_i$, by 
  $\len(h_i)$.
 Each SU has 
 its own required bandwidth and its own MAR.
 We denote the 
 required bandwidth of the $j$th user $u_j$,
 $1\leq j \leq N$,
 by $R_j$, and its MAR by $\delta_j$.

 The objective is to maximize the total spectrum utilization by assigning a subset of 
 the
 available spectrum holes to each SU.
 This assignment has to be carried out
 subject to the following conditions:
 \begin{itemize}
 	\item Each spectrum hole can be assigned
 	to at most one SU.
 	This means that it can be left unutilized.
 	\item The total length of the spectrum holes
 	assigned to an SU has to be greater than
 	or equal to its required bandwidth.
 	\item If $h_s=[\alpha_s,\beta_s]$
 	is the leftmost spectrum hole and
 	$h_e=[\alpha_e,\beta_e]$ is the rightmost
 	spectrum hole assigned to 
 	the SU $u_j$, $1\leq s\leq e\leq M$ and
 	$1\leq j\leq N$, then 
 	$\beta_e-\alpha_s$ cannot be greater than
 	$\delta_j$.
 \end{itemize}
 Therefore, a \textit{feasible hole assignment scheme
 	(pattern)}
 for the SU $u_j$, $1\leq j\leq N$,
 is a subset of spectrum holes
 whose total length is greater
 than or equal to 
 $u_j$'s
 required bandwidth $R_j$,
 and satisfies 
 $u_j$'s
 MAR constraint.
 Mathematically speaking,
 if
 $1\leq i_1 < i_2 < \cdots < i_\ell \leq M$
 is some (increasing) integer sequence (of indices),
 then the subset $\pi=\{h_{i_1},h_{i_2},\ldots,h_{i_\ell}\}$ of $\calh$
 is a feasible hole assignment scheme for
 the SU $u_j$
 if, firstly, $\sum_{h\in \pi} \len(h)\geq R_j$
 and, secondly, $\beta_{i_\ell}-\alpha_{i_1} \leq \delta_j$.
 For the SU $u_j$,
 we denote
 by $\Pi_j$
 the set of all of its feasible hole assignment schemes.
 Finally,
 by
 $\bfi_\pi(\cdot)$
 we denote the indicator function of the 
 hole assignment pattern $\pi$;
 i.e., $\bfi_\pi(h)=1$, if $h\in \pi$, and 
 $\bfi_\pi(h)=0$, otherwise.

  \begin{table}[tbh]
 	\centering
 	\caption{ A summary of notations used in this paper.}
 	\label{tab.notat}
 	\begin{tabular}{|l|p{5in}|}
 		\hline
 		$ a_{ij} $ & The binary 
 		indicator variable for the assignment of
 		the hole $ h_i $ to the SU $ u_j $\\
 		$ {\cal F} $ & A set
 		of
 		forbidden 
 		patterns\\
 		$ {\cal H} $ & The set of all available
 		spectrum holes\\
 		$ H $ & A subset of $ \cal H $\\
 		$ h_i $ & The $ i $th available spectrum hole\\
 		$ \bfi_{\pi}(\cdot) $ & The indicator function of the 
 		hole assignment pattern $\pi$\\
 		 		$ \len(h) $ &The length of the hole $ h $\\
 		 $ M $ & The size of the set $ \cal H $\\
 		$ N $ & The size of the set $ \cal U $\\
 		$ R_j $ & The 
 		required bandwidth of the SU $u_j$\\
 		 $ {\cal U} $ & The set of all SUs\\		
 		$ u_j $ & The $ j $th user	\\
 		$ x_{j,\pi} $ & A binary decision variable in the formulation (BILP), and a real-valued decision variable in
 		its LP relaxation
 		(PLP)\\
 		$ y_i $ & A non-negative real-valued decision variable in (DLP)\\
 		$ z_j $ & A non-negative real-valued decision variable in (DLP)\\
 		\hline
 		$ \alpha_i $ & The left endpoint of the spectrum hole $ h_i $\\
  		$ \beta_i $ & The right endpoint of the spectrum hole $ h_i $\\	
  		$ \delta $ & The MAR of a user $ u $\\	
 		$ \delta_j $ & The 
 		MAR of the SU $u_j$\\
 		$ \Pi_j $ & The set of all feasible hole assignment schemes for the SU $ u_j $ (with respect to $ \cal H $)\\
 		$ \Pi_{u,H} $ & The
set
of all the feasible hole assignment patterns of a user $u$
with respect to a subset $H$ of $ \cal H $\\
 		$ \varrho $ & The required bandwidth of a user $ u $\\

 		%$ \nu_h $ & A binary
 		%decision
 		%variable for the formulation in Subsection~\ref{subsect.sep1}\\

 		\hline
 	\end{tabular}
 \end{table}

% \section{A Novel Integer Linear Programming Approach
% 	to the \probcap}
% 
%  In this section, we first provide a novel binary integer linear programming
% formulation of the \acro{}, and then, describe a
% \bnp{} procedure that 
% can effectively solve
% the corresponding instances.

 \section{A Novel 0-1 ILP Formulation of the \acro{} and 
 an Overview of the
\bnp{} Framework}
 The following proposition presents a
 novel
 0-1 ILP formulation of the problem.	
 \begin{proposition}\label{prop1}
 	The \prob{}  can be formulated as the following binary integer linear program:
 	\begin{IEEEeqnarray}{l}
 	\text{(BILP) Maximize} \quad \sum_{j=1}^{N} \sum_{\pi \in \Pi_j}R_j x_{j,\pi}, \label{obj}\\
 	\text{subject to }  \nonumber\\
 	\sum_{\pi \in \Pi_j} x_{j,\pi} \leq 1, \quad \text{for}\ j=1,2,\ldots,N, \label{constr1}\\
 	\sum_{j=1}^{N} \sum_{\pi \in \Pi_j} \bfi_{\pi}(h_i)\ x_{j,\pi} \leq 1,\quad \text{for}\ i =1,2,\ldots, M,\label{constr2}\\
 	x_{j,\pi}\in\{0,1\},\quad\text{for}\ j=1,2,\ldots,N\ \text{and}\ \pi \in \Pi_j.\label{binary.constr}
 	\end{IEEEeqnarray}
 \end{proposition}
 
 \begin{proof}
 	For every 
 	$j$, $j=1,2,\ldots,N$, and every 
 	$\pi \in \Pi_j$, we associate a binary
 	indicator variable $x_{j,\pi}$ with 
 	the pair $(j,\pi)$.
 	This decision variable has the following
 	interpretation: $x_{j,\pi}$ is $1$ if 
 	the hole assignment scheme $\pi$
 	takes part in the solution, and is $0$ otherwise.
 	In fact, $x_{j,\pi}$
 	indicates that whether or not
 	the spectrum holes in $\pi$ are allocated to 
 	the SU $u_j$.
 	The inequalities (\ref{constr1}) ensure that
 	at most one hole assignment scheme can be selected
 	for each SU.
 	(In a solution to the formulation,
 	the left-hand-side of one such constraint being zero
 	indicates that 
 	none of the spectrum holes are assigned to
 	the corresponding SU.)
 	The inequalities (\ref{constr2}) guarantee
 	that each spectrum hole can be occupied by at most one SU.
 	(In a solution to the formulation,
 	the left-hand-side of one such constraint being zero
 	implies that 
 	the corresponding hole is unoccupied.)
 	It is now clear that,
 	in the presence of constraints (\ref{constr1}) and 
 	(\ref{constr2}),
 	the objective function
 	that has to be maximized is the one
 	given in (\ref{obj}).
 \end{proof}
%Therefore, one may not be able to 
%describe such a 0-1 ILP
%instance in an explicit manner
%(i.e., to consider all the decision variables in an explicit and simultaneous manner),
%let alone solve it.

 The above-described formulation contains $\sum_{j=1}^N|\Pi_j|$ binary decision variables and $N+M$ linear constraints.
 The number of  decision variables  can grow exponentially in the worst-case scenario.
 As mentioned earlier, this exponential growth of the number of decision variables,
 which renders the general-purpose ILP solvers useless,
 calls for the application of the branch-and-price technique.
Roughly speaking, \bnp{}
is essentially nothing more than an LP-relaxation-based
\bnb{} framework in which the LP instances
are solved using the column generation technique.

 \subsection{Basics of LP-relaxation-based \bnb{} procedures}
 
 In an LP-relaxation-based \bnb{} procedure,
 the search space of the problem is represented as a tree of live
 (a.k.a., open or active) nodes.
 The root of this tree
 corresponds to the original 0-1 ILP instance and each node
 corresponds to a subproblem.
 In fact, in each node, the search is restricted 
 to those solutions consistent (compatible) with it. 
 In each node, 
 we have to solve an LP instance,
 which is obtained by 
 relaxing (dropping) the integrality constraints
 from  the 
 associated 0-1 ILP instance.
 When the aim is to maximize an objective function,
 which is the case in our formulation,
 the optimal solution
 to this LP instance
 provides an upper bound to the
 optimal solution to  
 the 0-1 ILP instance corresponding to the node
 (because it has less restrictive constraints).
 At the root node,
 if we were lucky and the optimal solution 
 to the corresponding LP instance is integral,
 then the node is fathomed and this integral 
 optimal
 solution is returned.
 Otherwise, the algorithm proceeds by creating two
 (or more)
 children 
 nodes (subproblems) for the root. 
 To create the children nodes,
 one can select
 a decision variable
 whose 
 value in the optimal solution for the LP relaxation
 is non-integer.
 The variable is called as the branching variable.
 The children inherit all of the constrains of their parent.
 Furthermore, in one child the value of this branching
 variable is set to zero, and in the other, it
 is set to one.
 There are various \emph{branching variable selection strategies} recommended in the literature~\cite{bnb,applied}.
 The division process is repeated 
 according
 to a pre-specified \emph{node
 selection 
 strategy}
 until all of the subproblems are 
 fathomed 
 (pruned, conquered)~\cite{bnb,applied}.
 By pruning a node, we mean
 the exclusion of 
 all the nodes in the subtree rooted at it,
 from further consideration.
 In an LP-relaxation-based \bnb{} procedure,
 we fathom a node
 whenever one of the following cases occurs (see, e.g., \cite{applied}):
 \begin{itemize}
 	\item \textit{Pruning by integrality},
 	which occurs when the optimal solution
 	to the corresponding LP instance is integral. In this case, the value of this integral solution is compared to the value of the best integral solution found so far, which is usually called as the \textit{incumbent solution}.
 	(We have to keep track of the best solution found so far.) 
 	If this new-found integral solution is 
 	better than the incumbent solution, then it needs to be updated. 
 	\item \textit{Pruning by bound},
 	which occurs when 
 	the value of an optimal solution to the LP instance corresponding
 	to the node is not better than the value of the incumbent solution. Such a node can be prunded out
 	because it
 	cannot lead us to a better solution.
 	\item \textit{Pruning by infeasibility},
 	which occurs when
 	 the LP instance corresponding to the node is infeasible. 
 \end{itemize} 
 In the second case, the node is said to be nonpromising
 because of its bound, and in the third case,
 the node is said to be nonpromising because of its infeasibility.
 
 \subsection{ The \bnp{} Framework}
 
 Since the bounding strategy 
 in the above-described procedure is 
 based on solving an LP instance
 within each node of the search tree,
 the effective solution of these LP instances becomes
 of a crucial importance.
 As one such linear program
 can involve a huge number of decision variables, we need to embed a column generation
 subroutine into the \bnb{} framework.
 In fact, 
 the fundamental difficulty we encounter
 in this \bnb{} framework is that 
 the LP instances may have exponentially large
 number of decision variables (columns), therefore
 we have to resort to the column
 generation 
 approach
 to generate 
 columns in an on-the-fly fashion.
 In the terminology of the \bnp{} approach,
 the LP instance corresponding to
 a node in the search tree is commonly called 
 as the master problem. 
 At each node,
 in order to solve the LP instance,
 we start with a small number of 
 decision
 variables. This confined LP instance is 
 called
 restricted master problem. After solving this
 problem, we have to determine
 whether the current solution is optimal.
 If it is not, 
 the decision variables (columns) 
 to be added to the model
 must be identified. % on-the-fly
 The problem of generating such
 improving columns, if at least one exists,
 or otherwise declaring the optimality of the solution at hand,
 is often called as the
 pricing (sub-)problem.
 This
 procedure continues until 
 it is confirmed that
 no further improvement could be made.

 \section{The Column Generation Procedure and the Pricing Oracles}
 
 Now we are in a position to describe our
 column generation approach for solving
 the LP instances. 
 It can easily be seen that,
 the employment of
 the column generation technique 
 for solving a primal linear program
 can be viewed as
 the use of 
 the row generation technique
 for solving
 its dual linear program~\cite{dual-2,dual-1,dual-3,dual-4}.
 In fact,
 in the column generation technique for solving
 a linear program,
 for verifying the optimality of the
 solution at hand, 
 we have to solve a subproblem, called as the pricing problem.
 This is exactly equivalent to solving
 the 
 so-called separation problem for the dual linear program.
 It seems to us that 
 describing the
 row generation
 scheme 
 for solving the dual linear program
 may be more accessible to the reader.
 Therefore,
 in what follows,
 we consider the dual linear program,
 and describe
 two \textit{separation oracles}.
 These are in fact
  \textit{pricing oracles}
 for the primal linear program.
 We have to choose the
 suitable pricing oracle
 according to
 our branching strategy.
 The LP relaxation of the 0-1 ILP formulation
 given in Proposition~\ref{prop1}
 and its associated dual linear program are as follows:

 $$\begin{array}{l}
 \hline
 \mbox{Primal Linear Program} \\
 \hline
 \text{\textbf{(PLP)} Maximize} \quad \sum_{j=1}^{N} \sum_{\pi \in \Pi_j}R_j x_{j,\pi}, \\
 \text{subject to } \quad \nonumber\\
 \sum_{\pi \in \Pi_j} x_{j,\pi} \leq 1, \quad \text{for}\ j=1,2,\ldots,N, \\
 \sum_{j=1}^{N} \sum_{\pi \in \Pi_j} \bfi_{\pi}(h_i)\ x_{j,\pi} \leq 1, \quad \text{for}\ i =1,2,\ldots, M,\\
 0 \leq x_{j,\pi}\leq 1,\quad\text{for}\ j=1,2,\ldots,N\ \text{and}\ \pi \in \Pi_j.
 \\
 \hline
 \end{array}$$

 $$
 \begin{array}{l}
 \hline
 \mbox{Dual Linear Program}\\
 \hline
 \text{\textbf{(DLP)} Minimize}\quad\sum_{i=1}^{M} y_i + \sum_{j=1}^{N} z_j, \\
 \text{subject to}\\
 z_j + \sum_{i=1}^M \bfi_{\pi}(h_i)\ y_i \geq R_j,\\
 \hspace{1.2in}\text{for}\ j=1,2,\ldots,N\ \text{and}\ \pi \in \Pi_j, \\
 y_i \geq 0, \quad \text{for}\ i =1,2,\ldots, M,\\
 z_j \geq 0, \quad \text{for}\ j=1,2,\ldots,N.\\
 \hline
 \end{array}
 $$
 
 In the dual linear program, the number of constraints (rows) is prohibitively large.
 This difficulty can be overcome
 by an ad hoc incorporation of the 
 constraints into the LP problem instance
 (i.e., in an as-needed fashion).
 In the dual linear program,
 we start with an 
 initial (small or even empty) set of constraints.
 In order to determine whether
 the solution at hand is \textit{feasible}, an instance of
 the separation problem
 has to be solved.
 We try
 to identify the
 \textit{violated} 
 linear constraints
 to be included in the current set of 
 constraints.
 If such rows are found,
 they are appended to
 the current set of constraints,
 and 
 the resulting linear
 program
 is reoptimized.
 This procedure is
 repeated iteratively until 
 no further violated 
 linear constraints can
 be found.
 Therefore, 
 starting from the initial
 set of linear inequalities, the constraint
 set is progressively expanded by 
 incorporating the violated
 constraints.
 Let $\bfy^*=(y^*_i)_{i=1,2,\ldots,M}$
 and
 $\bfz^*=(z^*_j)_{j=1,2,\ldots,N}$
 be two vectors of nonnegative real components.
 In order to determine whether
 $(\bfy^*,\bfz^*)$
 is a feasible solution to
 (DLP), it suffices to decide
 whether there exists  some 
 $j\in\{1,2,\ldots,N\}$
 and some
 $\pi \in \Pi_j$ such that
 $ \sum_{i=1}^M \bfi_{\pi}(h_i)\ y^*_i < R_j - z^*_j$.
 Therefore,
 the separation problem 
 can be stated as follows:
%
%\begin{quote}
% 	\textbf{Instance:} 
% 	The sets $\calh$ and $\calu$,
% 	a
% 	nonnegative real vector 
% 	$\mathbf{y}^*=(y_i^*)_{i=1,2,\ldots,M}$, and 
% 	an integer $j\in\{1,2,\ldots,N\}$.\\
% 	\textbf{Task: }
% 	Find a pattern ${\pi\in\Pi_j}$  such that 
% 	$\sum_{i=1}^M \bfi_{\pi}(h_i)\ y^*_i$
% 	is minimum.
%\end{quote}

 \begin{quote}
	\textbf{Instance:} A set
	of spectrum holes $H$, a
	nonnegative real vector 
	$\mathbf{y}^*$ of the same size, and
	a user $u$ whose required bandwidth
	is $\varrho$ and whose MAR
	is $\delta$.\\
	\textbf{Task: }
	Let $\Pi_{u,H}$ be the
	set
	of all the feasible hole assignment patterns of $u$
	\textit{with respect to the set $H$}.
	Find a pattern $\pi\in\Pi_{u,H}$  such that 
	$\sum_{h\in H} \bfi_{\pi}(h)\ y^*_h$
	is minimum, where
	$ y^*_h $
	is the element of $ \mathbf{y}^* $
	corresponding to the hole $ h\in H $
\end{quote}

We call 
the above-stated separation
problem
as {SEP1}.
 Indeed,
 $(\bfy^*,\bfz^*)$
 is a feasible
 solution to
 (DLP) if and only if,
 for every $ u_j $,
 $j\in \{1,2,\ldots,N\}$,
 the value of
 an optimal solution
 to SEP1 
 over the set $ \Pi_j = \Pi_{u_j,{\cal H}} $
 is greater than
 or equal to
 $R_j - z^*_j $.

 However, there is still
 an issue here that must be
 addressed.
 As stated earlier,
 in the branch-and-bound tree,
 in the linear
 program associated with a non-root node,
 we have additional constraints
 that enforce that 
 some of the decision variables $x_{j,\pi}$
 take the value zero, and some of them
 take the value one.
 A constraint that assigns the value
 one to a decision variable
 $x_{j,\pi}$ is straightforward to deal
 with. All we need to do is to 
 exclude the SU $u_j$ 
 from the set $\cal U$ and
 treat the holes in $\pi$
 as occupied
 (the gain earned by this assignment is $R_j$).
 More accurately speaking,
 if the subproblem
 corresponding to a
 (non-root) node
 of the tree
 is constrained by
 the 
 equalities
 $x_{j_1,\pi_1}=x_{j_2,\pi_2}=\cdots=
 x_{j_t,\pi_t}=1$,
 where 
 $1\leq j_1 < j_2 < \cdots < j_t \leq N$
 and, 
 for each $\ell\in\{1,2,\ldots,t\}$,
 $\pi_{\ell}\in \Pi_{j_\ell}$,
 then in this subproblem,
 the set of
 spectrum holes
 is
 \begin{equation}\label{hset}
 \calh\setminus\bigcup_{\ell=1}^{t}\pi_{j_\ell}
 \end{equation}
 and the set of SUs is
 $\calu\setminus \{u_{j_1},u_{j_2},\ldots,u_{j_t}\}$.
 Notice that the patterns $\pi_1,\pi_2,\ldots,\pi_t$ are pairwise disjoint. Moreover, the gain earned by the
 selection of
 these SUs is $\sum_{\ell=1}^{t}R_{j_\ell}$.
 On the other hand,
 a
 constraint  that requires
 the value of 
 a decision variable $x_{j,\pi}$
 to be zero is not as convenient to deal with.
 This variable cannot be selected
 to enter the basis
 (i.e., become a basic variable).
 Therefore,
 corresponding to each
 (non-root)
  node in the
 tree, we may have a set
 of decision variables
 that aren't allowed to be selected
 to enter the basis.
 Accordingly,
 our separation oracle needs
 to be able to solve the following
 more general problem, which we
 call SEP2:
 \begin{quote}
 	\textbf{Instance:} A set
 	of spectrum holes $H$, a
 	nonnegative real vector 
 	$\mathbf{y}^*$ of the same size,
 	a user $u$ whose required bandwidth
 	is $\varrho$ and whose MAR
 	is $\delta$,
 	and a subset ${\cal F} \subseteq \Pi_{u,H}$
 	of forbidden patterns, where $\Pi_{u,H}$ is the
 	set
 	of all the feasible hole assignment patterns of $u$
 	with respect to the set $H$.\\
 	\textbf{Task: }
 	Find a pattern ${\pi\in\Pi_{u,H} \setminus {\cal F}}$  such that 
 	$\sum_{h\in H}\bfi_{\pi}(h)\ y^*_h$
 	is minimum, where
 	$ y^*_h $
 	is the element of $ \mathbf{y}^* $
 	corresponding to the hole $ h\in H $.
 \end{quote}

 \subsection{Solving the SEP1 problem}\label{subsect.sep1}%{A note on the branching scheme}
 As we will see in the next subsection,
the SEP2 problem can be solved by the use of
the dynamic programming paradigm~\cite[Chapter~15]{clrs}.
However, instead of the usual branching on the binary decision variables
$ x_{j,\pi} $, another branching
strategy can be used
so that
 the SEP1 problem, which is easier to deal with,
 is solved, instead of SEP2.
 As stated above,
fixing a decision
variable 
$ x_{j,\pi} $
equal to zero,
forbids
a
hole assignment
pattern for the SU
$ u_j $.
As we go deep
in the search tree,
the size of the set of
forbidden (excluded)
patterns $ \cal F $
gets larger, and
solving the problem 
SEP2 can
become more challenging.
Hence,
  instead of the usual branching on the variables $ x_{j,\pi} $, we try to branch on the 
  variables
  $ a_{ij} $ ($ 1\leq i \leq M $ and $ 1 \leq j \leq N $), which are 
  in fact
  \emph{hidden and implicit}. 
  Setting variable $ a_{ij} $ to 1 means that
   the hole $ h_i $ is assigned to the user $ u_j $, while
   setting $ a_{ij} $ to 0 means that $ h_i $ is not assigned to $ u_j$.
   This is exactly analogous
   to what has been described in
   \cite[Subsec.~13.4.1]{applied}
   for the GAP.
   
   In a node of the search tree,
   for solving the problem SEP1
   for the SU $ u_{j_1} $,
   $ 1\leq j_1 \leq N $,
   we
   firstly have to exclude
   all the already occupied holes
   (i.e., the holes for which we have
   $ a_{ij} = 1$)
   from the set of available
   holes.
   If a hole is occupied by
   the SU $ u_{j_1} $ itself,
   then the hole contributes
   in providing the
    required bandwidth $ R_{j_1} $.
    Therefore,
     $ \varrho $ has to
     be set
    equal to $ R_{j_1} $
    minus the sum of the lengths of 
    the holes occupied by $ u_{j_1} $
   itself.
   Furthermore,
   every hole $ h_i $
   for which $ a_{i{j_1}} = 0$
   has  to be excluded from the
   set of available holes.

   Now, if we are sure that
   the set $ H $ of remaining holes 
   do satisfy the
   MAR constraint
   for $ u_{j_1} $,
   then
   SEP1 reduces to an
   instance of the 
   Minimization
   Knapsack Problem (MinKP)~\cite[Subsec.~13.3.3]{knapsack}:
   $$
   \begin{array}{ll}
   	\text{Minimize}& \sum_{h\in H}y_h^*\nu_h,\\
   	\text{Subject to} & \sum_{h\in H}\len(h)\nu_h\geq \varrho,\\
   	&\nu_{h}\in\{0,1\},\quad  h\in H.
   \end{array}
   $$ 
   In the above
   description of the MinKP, the
    binary
    decision
     variable $ \nu_h = 1 $ 
     if the hole $ h\in H $ is assigned to the SU
     $ u_{j_1} $ and $ \nu_h = 0 $ otherwise.
   Moreover, an optimal solution
   for a given instance of the
   MinKP
   can readily be obtained
   from  an optimal solution for the following instance of
   the traditional 0-1 knapsack problem
   (KP)~\cite[Subsec.~13.3.3]{knapsack}:
   $$
   \begin{array}{ll}
	\text{Maximize}& \sum_{h\in H}y_h^*\xi_h,\\
	\text{Subject to} & \sum_{h\in H}\len(h)\xi_h\leq \sum_{h\in H}\len(h)-\varrho,\\
	&\xi_{h}\in\{0,1\},\quad  h\in H.
	  \end{array}
   $$
   Finally,
   this resulting
   instance of the 0-1 knapsack problem
   can effectively be solved
   e.g. 
   using the dynamic programming
   approach \cite[Sec.~2.3]{knapsack}.
%   The computational cost
%   of each call
%   to this DP procedure is
%   $ O(m(\sum_{i=1}^m\len(h_i)-\varrho))
%   =O(M()) $
However,
we do not know whether the set $ H $
satisfies the MAR constraint
for the SU $ u_{j_1} $.
 Hence, before every call to the
 knapsack oracle,
 we must be sure that
 the set of holes we 
 consider, do
 satisfy  
 the MAR constraint for 
 the SU $ u_{j_1} $.
There are two cases to consider:
\begin{itemize}
	\item If the set of holes
	that have already been assigned to
	$ u_{j_1} $
	is empty,
	then for finding
	a pattern 
	in $ \Pi_{u_{j_1},H} $
	that minimizes  
	the objective function
	$\sum_{h\in H} \bfi_{\pi}(h)\ y^*_h$,
	we proceed as follows.
	For every $ h\in H $,
	we consider
	$ h $ as the
	first contributing hole,
	and consider all the holes
	in $ H $
	that are not \textit{too far} from $ h $
	as the other available holes.
	These holes together do not violate the MAR constraint for $ u_{j_1} $. 
	The knapsack oracle has to be called
	for every such a set.
	A
	pattern corresponding
	to the
	minimum value returned by these calls is a
	pattern in
	$ \Pi_{u_{j_1},H} $
	that minimizes  
	the objective function
	$\sum_{h\in H} \bfi_{\pi}(h)\ y^*_h$.
	
	\item
	If the set of holes
	that
	have already been 
	assigned to
	$ u_{j_1} $ is not empty,
	then
	for every $ h\in H $ that
	does not appears after the first
	already assigned hole,
	we consider
	$ h $ as the
	first contributing hole.
	If 
	$ h $ and 
	the already 
	assigned
	holes 
	are 
	too far apart 
	from each other to
	satisfy the MAR
	constraint for $ u_{j_1} $,
	then they cannot
	provide 
	a pattern
	in $ \Pi_{u_{j_1},H} $.
	Hence,
	we don't call
	the knapsack oracle
	for them.
	Otherwise,
we consider
$ h $ as the
first contributing hole,
and consider all the holes
in $ H $
that are not too far from $ h $
as the other available holes.
The knapsack oracle has to be called
for every such a set.
A 
pattern corresponding
to the
minimum value returned by these calls is a
pattern in
 $ \Pi_{u_{j_1},H} $
 that minimizes  
 the objective function
 $\sum_{h\in H} \bfi_{\pi}(h)\ y^*_h$.
 
\end{itemize}

 The \bnp{}
 procedure 
 for solving the
 (BILP) instances
 that branches
 on the variables
 $ a_{ij} $, and
 employs the
 above-described 
 separation oracle,
 will be called
 \sepi{} hereinafter.

 \subsection{Solving the SEP2 problem }\label{subsect.sep2}

 We can also employ the
 standard branching scheme.
 At each node of the search tree,
 we branch on
 a variable $x_{j,\pi}$
 whose value is fractional.
 Therefore,
 the separation oracle
 has to
 solve the more
 challenging problem SEP2. 
 Algorithm~\ref{oraclealg}
 presents
 the pseudocode of
 the recursive procedure
 $\oracle$
 that can effectively solve
 the instances of
 the SEP2 problem.
 %For problem instances
 %for which we are somehow sure that
 %the depth of the
 %search tree
 %does not increase too
 %much,
 As the simulation results show,
 employing the standard branching 
 scheme and 
 $\oracle$
 within the \bnp{} framework
 can present a
 very good performance.
  The \bnp{}
 procedure 
 for solving the
 (BILP) instances
 that branches
 on the variables
 $ x_{j,\pi} $, and
 employs 
 $ \oracle $
 as the
 separation oracle,
 will be called
 \sepii{} hereinafter.
 In our implementation of
 the \sepii{}, we employed the
 \textit{most infeasible branching strategy},
 in which the variable whose
 (fractional) value
 is closest to $0.5$ is chosen
 to be branched on~\cite{bnb}.

 The inputs to the
 procedure $ \oracle $,
 which is in fact a
 memoized top-down DP
 algorithm%
 \footnote{For a thorough discussion of
 	dynamic programming
 	and memoization, the reader is referred to~\cite[Chapter~15]{clrs}.},
 are as follows:
 \textbf{1)}~A set of spectrum holes 
 $H=\{\widetilde{h}_1,\widetilde{h}_2,\ldots,\widetilde{h}_m\}$.
 (We used tilded symbols to distinguish the members of $ H $ from the holes in
 $ \cal H $.)
 \textbf{2)}~A nonnegative
 real vector $\mathbf{y}^*$
 whose length is equal to the size of
 $H$.
 \textbf{3)}~An index $\sigma\in\{1,2,\ldots,|H|\}$
 that specifies the 
 first available hole (i.e.,
 the first $\sigma-1$ holes are not allowed to
 be used).
In
 the top-level call we always have
 $\sigma=1$.
 \textbf{4)}~A required bandwidth $\varrho$.
 \textbf{5)}~An MAR $\delta$.
 \textbf{6)}~A boolean flag that indicates whether
 a hole has already been
 selected to be included in the output pattern.
 In
 the top-level call we always have
 $\mathit{active}=\mathsf{False}$.
  This input is actually used to ensure that the MAR constraint is not violated. If  $ \mathit{active}=\mathsf{True} $, then we have already assigned at least one hole to the user and we must make sure that we do not go too far from the hole(s) that have been assigned so far. If we move too far away from the first assigned hole, and the MAR constraint becomes violated, we lose the feasibility. If $\mathit{active}=\mathsf{False}$, we should not worry about the violation of the MAR constraint. In fact, as soon as $ \mathit{active}$ becomes $\mathsf{True}$, we have to be concerned about the violation of the MAR constraint. 
 \textbf{7)}~A set $\calf$ of forbidden patterns.
 (Each element of $\calf$ is a non-empty subset
 of $H$.)

 The procedure 
 $\oracle$
 returns
 a feasible hole assignment scheme
 $\pi\notin \calf$  
 corresponding to
 the given 
 bandwidth requirement $\varrho$
 and MAR $\delta$
 (with respect to the set $H$)
 for which  the summation
 $\sum_{h\in H} \bfi_{\pi}(h)\ y^*_h$
 is minimized.
 If no
 feasible hole assignment
 scheme exists for
 the given instance, it returns an empty
 set. 
 In a non-root node of the
 search tree,
 for the SU $u_j$, $1\leq j \leq N$,
 all the patterns $\pi \in \Pi_j$
 for which we have $x_{j,\pi}=0$
 should be included in a set $\calf$.
 If none of the patterns in $\Pi_j$
 are forbidden, i.e., none of the variables $x_{j,\pi}$,
 $\pi\in \Pi_j$, are set to be zero,
 then we set $\calf$ to be $\varnothing$.
 This is the case, for example, in the root node.
 Then, for obtaining the
 violated
 constraints   corresponding to $u_j$,
 or to conclude that no such constraint exists,
 we have to call  
 $$\oracle(H,\bfy^*,1,R_j,\delta_j,\mathsf{False},\calf),$$
 where $H$
 is as defined 
 in~(\ref{hset}).
 (In the root node
 we have $H=\calh$.)

 We now argue the
 correctness of the procedure $ \oracle $. 
 Depending on the values of the arguments
 $\sigma$, $\varrho$, and $\delta$,
 we have to distinguish three cases:
 \textbf{Case~I:} $\sigma = |H|$,
 i.e,
 the first  available
 hole is indeed the last member of $ H $;
 \textbf{Case~II:} $\sigma < |H|$ and 
 $\varrho \leq \len(\widetilde{h}_\sigma)$,
 i.e., the first available hole
 is not the last hole in $ H $ and it \textit{alone}
 can  provide the 
 required bandwidth $ \varrho $;
 and, finally,
 \textbf{Case~III:}
 $\sigma < m$ and 
 $\varrho > \len(\widetilde{h}_\sigma)$,
 i.e., the first available hole
 is not the last hole in $ H $ and it alone
 cannot  provide the 
 required bandwidth $ \varrho $.
 The procedure $\oracle$
 is made up of six parts;
 three of them are corresponding to
 the above three cases (Parts IV--VI),
 two of them
 are responsible for detecting infeasibility
 (Parts~I and II),
 and one (Part~III) is responsible for  the
 retrieval of the
 previously stored
 (cached)
  results.
 These six parts are as follows:
 \begin{itemize}
 	\item \textbf{Part~I.} (Lines~\ref{d<r.b}--\ref{d<r.e})
 	If the MAR is negative or is strictly less
 	than the bandwidth requirement,
 	then, obviously, 
 	the problem instance is infeasible.
 	Since
 	there is no
 	feasible hole assignment for 
 	such values of $\varrho$ and $\delta$,
 	the procedure returns an empty set.
 	\item \textbf{Part~II.} (Lines~\ref{infeas.b}--\ref{infeas.e})
 	If,
 	for the given values of $\varrho$ and $\delta$,
 	for every $\sigma\leq s_1 \leq s_2 \leq m$,
 	the subset $\{\widetilde{h}_{s_1}, \widetilde{h}_{s_1+1},\ldots, \widetilde{h}_{s_2}\}$
 	of $H$ is not a feasible hole assignment
 	scheme,
 	then the problem instance is infeasible.
 	In fact, 
 	for the given values of $\varrho$ and $\delta$,
 	if no set of 
 	\textit{consecutive}  spectrum holes of
 	$H$
 	can be utilized,
 	because either $\sum_{i=s_1}^{s_2}\len(\widetilde{h}_i)<\varrho$
 	or $\widetilde{\beta}_{s_2} - \widetilde{\alpha}_{s_1} > \delta$,
 	then certainly no subset of
 	$H$,
 	whose holes are not necessarily consecutive,
 	is 
 	utilizable.
 	\item \textbf{Part~III.} (Lines~\ref{save.b}--\ref{save.e})
 	The procedure 
 	\oracle{}
 	employs a globally accessible 
 	lookup
 	table $ M $
 	in which it stores
 	all the subproblem solutions.
 	Whenever we want to solve a subproblem, we first check that 
 	whether
 	$ M $ contains a solution. 
 	If a solution
 	has already been stored in
 	$ M $, then 
 	we have to simply retrieve and return the
 	stored result.
 	Otherwise, 
 	unless the subproblem is sufficiently small,
 	we solve
 	it by recursive call(s) to
 	$ \oracle $ itself.
 	It is important to
 	note that,
 	$ \oracle $
 	is in fact
 	a 
 	recursive
 	divide-and-conquer
 	(top-down)
 	procedure.
 	However, we have to
 	avoid solving a
 	subproblem more than once.
 	Therefore, we
 	maintain the table $ M $,
 	which is in fact
 		a container of key-value
 		pairs,
 		one pair for each subproblem.
 		In each key-value pair,
 		 subproblem parameters are
 		 used as
 		  the key, and the value is a 
 		   solution to this
 		   subproblem.%
 		   \footnote{In our C++
 		   	implementation of
 		   	\oracle{}, we utilized
 		   	\texttt{std::map} for this purpose.}
 	\item \textbf{Part~IV.} (Lines~\ref{s=m.b}--\ref{s=m.e})
 	If the first available hole is
 	the $m$th
 	(i.e., the last) one, then the problem instance
 	is feasible only if
 	the following three conditions are 
 	satisfied:
 	$\varrho \leq \widetilde{\beta}_m-\widetilde{\alpha}_m$, $\delta\geq \widetilde{\beta}_m-\widetilde{\alpha}_m$, and
 	$\{\widetilde{h}_m\}\not\in\calf$. In this case,
 	the procedure returns the 
 	only feasible solution that is $\{m\}$.
 	Otherwise, no feasible
 	hole assignment scheme exists
 	for the given values of $\sigma$,
 	$\varrho$, $\delta$, and $\calf$. Therefore,
 	the procedure returns an empty set.
 	\item \textbf{Part~V.} (Lines~\ref{r<first.b}--\ref{r<first.e})
 	If $\sigma < m$ and $\varrho \leq \len(\widetilde{h}_\sigma)$,
 	then
 	we have to consider four subcases.
 	If
 	$\delta < \len(\widetilde{h}_\sigma)$
 	and $\mathit{active}=\mathsf{True}$,
 	then the instance is
 	simply infeasible, so the algorithm
 	returns an empty set.
 	If $\delta < \len(\widetilde{h}_\sigma)$
 	and $\mathit{active}=\mathsf{False}$,
 	then 
 	because of the 
 	violation of the
 	MAR constraint,
 	$\{\widetilde{h}_\sigma\}$
 	cannot be a feasible
 	hole assignment pattern.
 	Therefore,
 	the procedure should
 	call
 	$$\oracle(H,\boldy^*,\sigma+1,\varrho,\delta,\myactive,\calf')$$ and return the resulting set, where
 	$\calf'=\{F\in\calf\,\vert\,
 	F\subseteq \{\widetilde{h}_{\sigma+1},\widetilde{h}_{\sigma+2},\ldots,\widetilde{h}_{m}\}\}$.
 	If $\delta \geq \len(\widetilde{h}_\sigma)$
 	and $\mathit{active}=\mathsf{True}$,
 	then either we utilize
 	the spectrum hole
 	$\widetilde{h}_\sigma$ or we don't utilize it.
 	In the
 	former case,
 	the
 	hole assignment 
 	pattern
 	is $\{\widetilde{h}_\sigma\}$,
 	and in the latter
 	case,
 	the hole assignment pattern
 	is
 	the one
 	corresponding
 	to the set, call it $U$,
 	returned by the call
 	\begin{equation*}
 		\oracle(H,\boldy^*,\sigma+1,
 		\varrho,\delta-(\widetilde{\alpha}_{\sigma+1}-\widetilde{\alpha}_\sigma),\mathit{active},\calf').
 	\end{equation*}
 	Therefore,
 	if $\{\widetilde{h}_\sigma\}\in \calf$
 	or
 	$\sum_{u\in U}y^*_u<y^*_\sigma$,
 	the procedure should return the
 	set $U$,
 	and otherwise it should return
 	the set
 	$\{\sigma\}$.
 	It should be remarked that,
 	throughout the pseudocode,
 	our convention is that a summation over an empty index set  is considered to be $+\infty$.
 	Finally, if $\delta \geq \len(\widetilde{h}_\sigma)$
 	and $\mathit{active}=\mathsf{False}$,
 	then,
 	again,  either  we utilize
 	the spectrum hole
 	$\widetilde{h}_{\sigma}$ or we don't
 	utilize it.
 	In the former case,
 	the hole assignment pattern is
 	$\{\widetilde{h}_\sigma\}$, and
 	in the latter case,
 	the hole assignment pattern is
 	the one corresponding
 	to the set, call it $U$,
 	returned by the call
 	$$\oracle(H,\boldy^*,\sigma+1,\varrho,\delta,\myactive,\calf').$$
 	Note that because $\mathit{active}=\mathsf{False}$,
 	i.e., no hole has been assigned to the user so far, $ \delta $ remains unchanged. Unlike when $ \mathit{active} = \mathsf{True} $. 
 	Again,
 	if $\{\widetilde{h}_\sigma\}\in \calf$
 	or
 	$\sum_{u\in U}y^*_u<y^*_\sigma$,
 	the procedure should return the
 	set $U$,
 	and otherwise it should return
 	the set
 	$\{\sigma\}$.

 	\item \textbf{Part~VI.} (Lines~\ref{fin.b}--\ref{fin.e})
 	If $\sigma < m$ and $\varrho > \len(\widetilde{h}_\sigma)$,
 	then either the spectrum hole
 	$\widetilde{h}_\sigma$ takes part in an optimal solution
 	to the problem instance or does not.
 	In the case that $\widetilde{h}_\sigma$ does not contribute,
 	if $\mathit{active}=\mathsf{False}$,
 	then the 
 	optimal
 	solution is
 	the one corresponding to the set returned by the call
 	$$\oracle(H,\boldy^*,\sigma+1,\varrho,\delta,\myactive,\calf'),$$
 	and if $\mathit{active}=\mathsf{True}$,
 	then the 
 	optimal solution
 	is
 	the one corresponding to the set returned by the call
 	\begin{equation*}
 		\oracle(H,\boldy^*,\sigma+1,\\
 		\varrho,\delta-(\widetilde{\alpha}_{\sigma+1}-\widetilde{\alpha}_\sigma),\myactive,\calf').
 	\end{equation*}
 	In either case, we call the returned set $U$.
 	In the case that $\widetilde{h}_\sigma$	 contributes,
 	then the call
 	\begin{equation*}
 		\oracle(H,\boldy^*,\sigma+1,\\
 		\varrho - (\widetilde{\beta}_\sigma - \widetilde{\alpha}_\sigma),\delta-(\widetilde{\alpha}_{\sigma+1}-\widetilde{\alpha}_\sigma),\mathsf{True}, \calf'')
 	\end{equation*}
 	needs to be made, where
 	\begin{equation*}\calf''=\{F\setminus\{\widetilde{h}_\sigma\}\,\vert\,F\in\calf \ \mbox{and}\ 
 		F\subseteq \\ \{\widetilde{h}_{\sigma},\widetilde{h}_{\sigma+1},\ldots,\widetilde{h}_{m}\}\  
 		\mbox{and}\ \widetilde{h}_\sigma\in F\ 
 		\mbox{and}\ |F|>1\}.
 	\end{equation*}
 	Let us call the returned set $V$.
 	If
 	$\sum_{u\in U}y^*_u<
 	\sum_{v\in V}y^*_v+
 	y^*_\sigma$,
 	the procedure should return the
 	set $U$,
 	and otherwise it should return
 	the set
 	$V\cup\{\sigma\}$.
 \end{itemize}%

%Because of its recursive
%nature,
%establishing 
%a tight upper
%bound on the
%worst-case
% time complexity
%of $ \oracle $
%is a challenging
%task.
%We don't know 
%how many subproblems
%need to be solved
%for
%solving the original
%problem instance.

% A loose 
%and pessimistic
A worst-case
time complexity
for the procedure
\oracle{}
can be obtained
by 
multiplying together
the number of 
all possible
values to the
parameters
$ \sigma $,
$ \varrho $,
$ \delta $,
$\mathit{active}$,
and
$ \mathcal{F} $.
This is in fact
a straightforward application of the rule of product for counting the number of
all possible subproblems.
For instance, it can readily be
seen that,
if $ {\cal F}=\varnothing $, then
the
worst-case time complexity of this procedure is in 
$\mathcal{O}\left(\frac{|H|\,\cdot\, \varrho \,\cdot\, \delta}%
{\min_{h\in H}{\len(h)}\,\cdot\,
	\min_{1\leq i \leq |H|-1}%
	(\widetilde{\alpha}_{i+1}-\widetilde{\alpha}_{i})}\right)$.
Under the assumption that
all $\widetilde{\alpha}_i$'s
and all $\widetilde{\beta}_i$'s are integers,
$1\leq i \leq |H|$,
this is indeed $\mathcal{O}(|H|\cdot\varrho\cdot \delta)$.
%%
%(This assumption does not affect the generality
%of the problem \cite{ilp}.)
However, 
the number of 
subproblems
actually needed to
be solved 
is far less than the 
above-mentioned upper bound,
because the procedure
only solves
the definitely required
subproblems~\cite[Sec.~15.3]{clrs}.
%This is
%intrinsic
%to the memoized version
%of recursive procedsures.

 We conclude this section by giving a tiny example,
 that may shed some light on the
whole process of \sepii{}.
 
 \begin{algorithm*}
 	\singlespacing\footnotesize
 	\caption{The memoized top-down dynamic programming
 		algorithm $\oracle$ for 
 		solving the separation problem SEP2.}\label{memoized}
 	%\small
 	\begin{algorithmic}[1]
 		\Statex \textbf{Input:}
 		\textbf{1)}~A set of spectrum holes $H$;
 		\textbf{2)}~A nonnegative
 		real vector $\mathbf{y}^*$
 		whose length is equal to the size of
 		$H$;
 		\textbf{3)}~An index $\sigma\in\{1,2,\ldots,|H|\}$
 		that specifies the 
 		first available hole;
 		\textbf{4)}~A required bandwidth $\varrho$;
 		\textbf{5)}~An MAR $\delta$;
 		\textbf{6)}~A boolean flag that indicates whether
 		a hole has already been
 		selected to be included in the output pattern;
 		\textbf{7)}~A set $\calf$ of forbidden patterns.
 		\Statex \textbf{Output:}
 		A subset $I$ of
 		$\{\sigma,\sigma+1,\ldots,|H|\label{key}\}$
 		for which $\sum_{i\in I}y^*_i$ is minimized.
 		(Throughout the pseudocode,
 		our convention is that a summation over an empty index set  is considered to be $+\infty$.)
 		\Statex
 		\textbf{Note:}
 		 The procedure uses a globally accessible 
 		 lookup
 		 table $ M $
 that maintains
all the subproblem solutions.
Initially,
$ M $ is empty.
 		\Procedure{\oracle}{$H,\boldy^*,\sigma,\varrho,\delta,\myactive,\calf$}
 		\If{$\delta < 0$ {\bf or} $\delta < \varrho$} \label{d<r.b} \Comment{Infeasible instance}
 		\State \bfreturn{} $\varnothing$;
 		\EndIf \label{d<r.e}
 		\State Let $m=|H|$; \Comment{$m$ is the number of spectrum holes in $H$, which is equal to the length of $\bfy^*$}
 		\If{for every $\sigma\leq s_1 \leq s_2\leq m$, we have
 			$\sum_{i=s_1}^{s_2}(\widetilde{\beta}_i-\widetilde{\alpha}_i) < \varrho$ or
 			$\widetilde{\beta}_{s_2}-\widetilde{\alpha}_{s_1}>\delta$} \Comment{Infeasible instance} \label{infeas.b}
 		\State \bfreturn{} $\varnothing$;
 		\EndIf \label{infeas.e}
 		\If{$ M $
 			contains 
 			a solution to the input instance} \label{save.b}
 		%\Comment{The retrieval of the already stored results}
 		\State retrieve and return the stored result;
 		\EndIf \label{save.e}
 		\If{$\sigma = m$} \Comment{A base condition} \label{s=m.b}
 		\If{$\varrho \leq \widetilde{\beta}_m-\widetilde{\alpha}_m$ \bfand{} $\delta\geq \widetilde{\beta}_m-\widetilde{\alpha}_m$ \bfand{}
 			$\{\widetilde{h}_m\}\not\in\calf$}
 		\State \bfreturn{} $\{m\}$;
 		\Else \State{\bfreturn{} $\varnothing$};
 		\EndIf
 		\EndIf \Comment{In the remaining part
 			of the algorithm, we have $\sigma < m$} \label{s=m.e}
 		\State $\calf'=\{F\in\calf\,\vert\,
 		F\subseteq \{\widetilde{h}_{\sigma+1},\widetilde{h}_{\sigma+2},\ldots,\widetilde{h}_{m}\}\}$;
 		\If{$\varrho\leq \widetilde{\beta}_\sigma-\widetilde{\alpha}_\sigma$} \label{r<first.b}
 		\If{$\delta<\widetilde{\beta}_\sigma-\widetilde{\alpha}_\sigma$ \bfand{} $\mathit{active} = \mathsf{True}$} \Comment{A base condition}
 		\State \bfreturn{} $\varnothing$;
 		\ElsIf{$\delta<\widetilde{\beta}_\sigma-\widetilde{\alpha}_\sigma$ \bfand{} $\mathit{active} = \mathsf{False}$}
 		\State \bfsr{} \oracle($H,\boldy^*,\sigma+1,\varrho,\delta,\myactive,\calf'$);\Comment{Stores the solution in $ M $}
 		\ElsIf{$\delta\geq\widetilde{\beta}_\sigma-\widetilde{\alpha}_\sigma$ \bfand{} $\mathit{active} = \mathsf{True}$}
 		\State $U=$ \oracle($H,\boldy^*,\sigma+1,\varrho,\delta-(\widetilde{\alpha}_{\sigma+1}-\widetilde{\alpha}_\sigma),\myactive,\calf'$);
 		\ElsIf{$\delta\geq\widetilde{\beta}_\sigma-\widetilde{\alpha}_\sigma$ \bfand{} $\mathit{active} = \mathsf{False}$}
 		\State $U=$ \oracle($H,\boldy^*,\sigma+1,\varrho,\delta,\myactive,\calf'$);
 		\EndIf
 		\If{$\{\widetilde{h}_\sigma\}\in\calf$ \textbf{or} $y_\sigma^*> \sum_{u\in U}y^*_u $}
 		\State \bfsr{} $U$;\Comment{Stores the solution in $ M $}
 		\Else
 		\State \bfsr{} $\{\sigma\}$;\Comment{Stores the solution in $ M $}
 		\EndIf 
 		\EndIf \label{r<first.e} \Comment{In the remaining part
 			of the algorithm, we have $\sigma < m$  and $\varrho> \widetilde{\beta}_\sigma-\widetilde{\alpha}_\sigma$}
 		\If{$\mathit{active} = \mathsf{True}$} \label{fin.b}
 		\State $U = $ \oracle($H,\boldy^*,\sigma+1,\varrho,\delta-(\widetilde{\alpha}_{\sigma+1}-\widetilde{\alpha}_\sigma),\myactive,\calf'$);
 		\Else
 		\State $U = $ \oracle($H,\boldy^*,\sigma+1,\varrho,\delta,\myactive,\calf'$);
 		\EndIf
 		\State $\calf''=\{F\setminus\{\widetilde{h}_\sigma\}\,\vert\,F\in\calf \ \mbox{and}\ 
 		F\subseteq \{\widetilde{h}_{\sigma},\widetilde{h}_{\sigma+1},\ldots,\widetilde{h}_{m}\}\  
 		\mbox{and}\ \widetilde{h}_\sigma\in F\ 
 		\mbox{and}\ |F|>1\}$;
 		\State $V=$ \oracle($H,\boldy^*,\sigma+1,\varrho - (\widetilde{\beta}_\sigma - \widetilde{\alpha}_\sigma),\delta-(\widetilde{\alpha}_{\sigma+1}-\widetilde{\alpha}_\sigma),\mathsf{True}, \calf''$);
 		\If{$\sum_{u\in U}y^*_u\leq \sum_{v\in V}y^*_v + y^*_\sigma$}
 		\State \bfsr{} $U$;\Comment{Stores the solution in $ M $}
 		\Else
 		\State \bfsr{} $V\cup \{\sigma\}$;\Comment{Stores the solution in $ M $} 
 		\EndIf \label{fin.e}
 		
 		\EndProcedure	
 	\end{algorithmic}
 	\label{oraclealg}
 	
 \end{algorithm*}

\begin{figure}[htb]\centering
 	\includegraphics[width=0.45\textwidth]{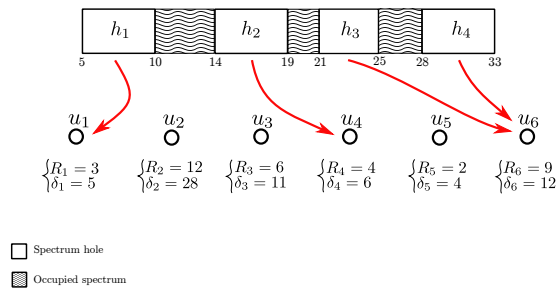}
 	\caption{A depiction of an optimal solution 
 		to the tiny instance 
 		described in Example~\ref{example1}.}
 	\label{fig}
 \end{figure}

 \begin{example}\label{example1}
 	Consider an instance of the problem in which
 	the set $\calh$ of available spectrum holes
 	consists of the following 4 holes
 	$$h_1= 	[5,10],
 	h_2= 	[14,19],
 	h_3= 	[21,25],
 	h_4= 	[28,33],$$
 	and the set $\calu$ of SUs includes the following 6 users:
 	$$
 	\begin{array}{ll}
 	u_1: 	       R_1=3 ,\,    \delta_1= 5,&
 	u_2: 	      R_2=12 ,\,    \delta_2= 28,\\
 	u_3: 	       R_3=6 ,\,    \delta_3=  11,&
 	u_4: 	       R_4=4 ,\,    \delta_4= 6,\\
 	u_5: 	       R_5=2 ,\,    \delta_5=  4,&
 	u_6: 	       R_6=9 ,\,    \delta_6=  12.
 	\end{array}
 	$$
 	In the root node of the search tree,
 	the column generation procedure
 	has to solve a linear program
 	in which the number of 
 	(functional) constraints is 10.
 	(See the linear program (PLP).)
 	In the first iteration of the procedure,
 	the objective function coefficients
 	are all zero.
 	The optimal value of this 
 	objective function is obviously zero.
 	In the second iteration,
 	we call the pricing algorithm
 	once
 	for each SU.
 	Each of these calls provides an improving
 	decision variable.
 	The objective function
 	(to be maximized)
 	will be therefore
 	\begin{equation*}
 	\varphi_1=3\,x_{1 , \{h_4\}} +
 	12\,x_{2 , \{h_2 , h_3 , h_4\}} +
 	6\,x_{3 , \{h_2 , h_3\}} +
 	4\,x_{4 , \{h_4\}} +
 	2\,x_{5 , \{h_3\}} +
 	9\,x_{6 , \{h_3 , h_4\}}.
 	\end{equation*}
 	In an
 	optimal solution to this linear program,
 	$x_{2 , \{h_2 , h_3 , h_4\}}=1$
 	and all other decision variables are zero.
 	The value of this solution is 12.
 	In the third iteration,
 	by calling the pricing algorithm once for each SU,
 	we find that there exist 4 improving decision
 	variables.
 	The objective function will be
 	\begin{equation*}
 	\varphi_2=\varphi_1+3\,x_{1 , \{h_2\}}  +
 	12\,x_{2 , \{h_1, h_2, h_4\}}+\\
 	4\,x_{4 , \{h_2\}}  +
 	9\,x_{6 , \{h_2 , h_3\}}.
 	\end{equation*}
 	In an optimal solution to this linear program,
 	$
 	x_{6 , \{h_3 , h_4\}}=
 	x_{2 , \{h_1, h_2, h_4\}}=
 	x_{6 , \{h_2 , h_3\}}=\frac{1}{2}$
 	and all other decision variables are zero.
 	The value of this solution is 15.
 	In the forth iteration,
 	we again run the pricing algorithm
 	once for each SU.
 	This indicates that there exist
 	3 improving decision variables.
 	The objective function will be
 	$$\varphi_3=\varphi_2+3\,x_{1 , \{h_1\}} + 
 	12\,x_{2 , \{h_1 , h_3 , h_4\}} + 
 	4\,x_{4 , \{h_1\}}.$$
 	In an optimal solution to this
 	linear program,
 	$
 	x_{6 , \{h_3 , h_4\}}=
 	x_{2 , \{h_1, h_2, h_4\}}=
 	x_{6 , \{h_2 , h_3\}}=
 	x_{4 , \{h_1\}}=\frac{1}{2}
 	$
 	and all other decision variables are zero.
 	The value of this solution is 17.
 	In the fifth iteration,
 	the pricing algorithm 
 	implies that there exists no improving
 	decision variable.
 	Therefore,
 	no further improvement can be made, and
 	the last obtained solution is 
 	optimal for the relaxed problem.
 	However, this optimal solution is non-integral, and
 	a variable (with a non-integer value)
 	has to be selected to be branched on.
 	(We will shortly see that the absolute integrality gap is in fact 1.)
 	We select the variable
 	$x_{6 , \{h_3 , h_4\}}$
 	(whose value is $0.5$ in the optimal solution)
 	as the branching variable.
 	The root node is split into two child nodes.
 	In one of them, we set the value of $x_{6 , \{h_3 , h_4\}}$ 
 	to be 1, and in the other, 0.
 	In the node in which
 	$x_{6 , \{h_3 , h_4\}}=1$,
 	the column generation procedure
 	returns an integral solution whose objective
 	value is 16. In this solution, we have
 	$x_{6 , \{h_3 , h_4\}} = 1$, $x_{1,\{h_1\}}=1$,
 	and $x_{4,\{h_2\}}=1$.
 	The value of all other decision variables is zero.
 	Since the solution is integral, we can fathom the
 	node.
 	On the other hand, in the node in
 	which 
 	$x_{6 , \{h_3 , h_4\}}=0$,
 	the column generation procedure 
 	converges to a non-integral solution
 	for which the value of the objective function
 	is
 	$16.75$.
 	In the considered instance of the problem,
 	all the bandwidth requirements are integral.
 	Hence,
 	the value of an optimal solution to it, has
 	to be integral as well.
 	Therefore, the current node cannot lead us
 	to a solution whose objective value is better that
 	16, and can safely be pruned.
 	This means that the solution found in the
 	first considered child node,
 	whose value was 16, is in fact
 	the optimal solution to the instance.
 	In this solution,
 	which is depicted in
 	Figure~\ref{fig},
 	$h_1$ is assigned to $u_1$,
 	$h_2$ is assigned to $u_4$, and
 	$h_3$ and $h_4$ are assigned to $u_6$.
 \end{example}
 \section{Computational Results}\label{simulsec}

 This section is devoted to evaluating the performance of
 the proposed 
 \bnp{} procedure
 against the best currently available
 ILP formulation
 of the problem, which is
 presented
 in~\cite{ilp},
 with respect to the CPU time.
 We implemented the \bnp{} procedure
 in C++, and 
 carried out 
 the simulations
 on an Intel$^{\text{\textregistered}}$ Core\texttrademark{}~i7-9750H 2.60~GHz laptop with
 16.00~GB~of~RAM, 
 running Microsoft$^{\text{\textregistered}}$ Windows$^{\text{\textregistered}}$~10 (64-bit) operating system.
 In our implementation of the
 \bnp{} procedure, for solving the LP instances
 in the nodes of the search tree,
 we employed the IBM$^{\text{\textregistered}}$ ILOG$^{\text{\textregistered}}$ Concert CPLEX$^{\text{\textregistered}}$ API. %(Ver.~12.10.0.0).

 As pointed out in the introduction, 
 two 0-1 ILP formulations 
 have been proposed in~\cite{ilp}
 for the problem. 
 The 
 (binary) integer
 linear programs
 obtained
 by using these formulations
 can be solved
 using the off-the-shelf ILP solvers like
 CPLEX and GUROBI.
 The second formulation, entitled 
 ``Linear Assignment Model (Type 2)," outperforms
 the first one, in terms of running
 time, and therefore, comparisons
 will be conducted against this formulation.
 Hereinafter we use the acronym \their{} to refer to it.
 For the sake of 
 self-containedness,
 we prsenet this formulation here.
 For every $j\in\{1,2,\ldots,N\}$
  and every
  $  i \in \{1,2,\ldots,M\} $, let
  the set $I_j(i)$ be defined by 
  $
  I_j(i)=\{i'\in\{i,i+1,\ldots,M\}\,|\, \beta_{i'}-\alpha_{i}\leq \delta_j \}
  $.
  Moreover,
  for every $j\in\{1,2,\ldots,N\}$, let
  the set $T_j$ 
  be defined as 
  $T_j=\{i\in\{1,2,\ldots,M\}\,|\,  \sum_{i'\in I_j(i)} \len(h_{i'})\geq R_j \}$.
  Finally, let the set $ \Gamma $ be defined as
  $
  \Gamma = 
  \{(j,i,i')\,|\, j\in\{1,2,\ldots,N\},\, i \in T_j ,\,  i' \in I_j(i)\}
  $.
 Corresponding to each triple $(j,i,i')\in \Gamma $,
 the formulation \their{}
 contains a binary 
 decision  variable  $ \gamma^{j}_{ii'} $.
 The formulation is as follows:
  	\begin{IEEEeqnarray*}{l}
 	\text{(\their{}) Maximize} \quad \sum_{j=1}^{N} R_j
 	\sum_{i\in T_j}\gamma^{j}_{ii}\ , \\
 	\text{subject to }\\
 	\sum_{(j,i,i')\in \Gamma} \gamma_{ii'}^j\leq 1,\quad\text{for}\  i'=1,2,\ldots,M,\\
 	\sum_{i\in T_j} \gamma_{ii}^j\leq 1, \quad \text{for}\ j=1,2,\ldots,N,\\
 	\sum_{i'\in I_j(i)}\len(h_{i'})\cdot\gamma^j_{ii'}\geq R_j \cdot \gamma^j_{ii}\ ,\quad \text{for}\  j= 1,2,\ldots,N
 	\ \text{and}\ 
 	 i\in T_j\ , \\
 	\gamma_{ii'}^{j}\in\{0,1\},\quad\text{for every}\ (j,i,i')\in\Gamma.
 \end{IEEEeqnarray*}	
 
 \their{}
 consists of at most $NM^2$ binary decision variables
 and at most $M + N + NM$ linear constraints.
 In contrast to our formulation (BILP) which,
 due to its huge number
 of decision variables,
 cannot be fed directly to an ILP solver,
 the integer linear programs  corresponding to \their{},
 due to their compactness,
 can be solved by using an ILP solver.
 In our experiments, the CPLEX solver
 (Ver.~12.10.0.0)
 has been employed,
 both as an LP solver within the \bnp{} routine and 
 as an ILP solver for solving the integer linear programs 
 corresponding to
 \their{}.
 
 There are two remarks that should be made at this point.
 Firstly,
 the
 performance of the
 \bnp{} search procedure
 can be improved 
 by embedding a 
 heuristic method
 for finding 
 good
 integer-feasible
 solutions during
 the search process~\cite{bnb}.
 We didn't employ 
 such a 
 heuristic
 method in the \bnp{} process.
 However, by means of a
 powerful heuristic
 method
 for generating good
 integer-feasible solutions,
 the active
 nodes
 can more effectively be fathomed,
 which in turn makes the search
 process faster.
 Secondly,
 a decision 
 has to be made as to
 which active node to visit
 first.
 Various node selection strategies
 have 
 been described in the literature~\cite{bnb}.
 In our implementation of
 the \bnp{} procedure, 
 we employed the
 \textit{best-bound-first} node
 selection rule.
 The unfathomed subproblems are maintained in a priority queue,
 and
 the one with the largest 
 LP relaxation bound
 has the highest priority.
 For the case of two nodes
 with equal LP relaxation bounds,
 the one that is more likely
 to yield an integer-feasible 
 solution is chosen.

 To our knowledge, 
 presently no  benchmark
 dataset is available   for the \acro{}.
 We therefore created our own one, which contains
 more than 1000
 randomly generated
 instances of the problem. 
 The
 comparison has been made 
 by the use of these
 instances.
% 100 instances
% for each considered value of 
% $M$ and $N$. 
 %
 All the problem instances, and also their optimal solutions,
 are available online.%
 \footnote{\url{https://github.com/hfalsafain/crrap}}%
 \footnote{At this web address, we have also provided a number of large instances of the problem that \sepi{}
 	and \sepii{}
 	 were unable to solve in less than 15 minutes.
  These instances may be
useful for future research on the problem.}
 %through the URL
 As in \cite{ilp}, 
  the TV band
  (470MHz -- 862MHz)
has been considered for the numerical simulations.
In all the generated 
instances,
the required bandwidth
of a user is a random number
(not necessarily integer)
  drawn
  uniformly
   from the
range $ [10,25] $ (in MHz).
This is similar to that of
\cite{ilp},
except that
in our instances,
the numbers are not necessarily integers.
%
%To generate a random
%instance,
%in addition to more
%basic parameters
%such as
%$ |\cal H| $ and
%$ |\cal U| $,
%we can specify the percentage of available frequency.
In the tables presented
in this section, 
the parameter $ q\in [0, 1] $ 
determines the
fraction
of available
spectrum 
in the frequency
band, which is mainly
determined by the 
geographical location.
It has been shown that in urban areas,
we have
$ q \approx 0.2 $, whereas  in suburban areas,
$ q \approx 0.8 $~\cite{ilp}.
A time limit of 600 seconds has been set for each run.
If the execution time for
a run
 reaches this limit, 
 the run halts and
the best-so-far 
solution is
 reported. 

A comparison of the average running times (in seconds)
of 
\sepi{}
and \sepii{}
 against
those of the formulation \their{},
is tabulated in Table~\ref{tab1-h25}.
%We implemented the \bnp{}
%approach based on
%both of the branching approaches
%described
%in the previous
%section.
The results
(optimal objective values)
 of
the three methods
were entirely consistent with each other.
For each row, we 
have considered 20 randomly generated instances of the problem. 
In the
upper part of the table
(first eight rows), as in \cite[Table~1]{ilp},
for a user $ u_j $,
the parameter $ \delta_j $
is drawn  uniformly
from the 
interval $ [2R_j,3R_j] $.
Again, unlike
the scenario
described in \cite{ilp},
$ \delta_j $s are 
not necessarily integers.
In this part, $ q $ is set to $ 0.5 $. Moreover, the number of 
available
holes for all 
the considered
instances is equal to $ 25 $, but for the number of users, we have 
$ N\in\{25,50,\ldots,200\} $.
In the lower part of the 
table (last six rows),
we have 
$ M=30 $,
$ N\in\{30,60,\ldots,180\} $,
and
$ q=0.25 $.
All $ \delta_j $s are
set equal to 45
$(1\leq j \leq N)$.
As can be observed from the table, when the number of users is small, there is no major difference between the running times,
but when the number of users is large, our approach has a much better
performance.

      \begin{table}[tbh]
	\centering
	\caption{The effect of the ratio of the number of users to the number of holes on the 
		time
		performance 
		(in seconds)
		of the approaches. For each row, 20 random instances are used.
		The column labeled with ``TL'' reports the number of 
		CPLEX
		runs 
		stopped due to the time limit
		(for solving \their{} models). None of the \sepi{} and \sepii{} runs reached the 600 seconds time limit.}
	\label{tab1-h25}
	\small
	\begin{tabular}{|l|c|c|c|cc|cc|}
		\hline
		$ M $ & $ N $ & $ q $  & $ \delta $ & \sepi{} & \sepii{} & \their{} & TL \\
		\hline
		25 & 25 & 0.5 & As in \cite[Table~1]{ilp} &1.34 &0.705 & \textbf{0.212} & 0\\ 	
		  & 50 &  &   &1.30 &\textbf{0.670} & 0.795 & 0\\
		  & 75 &   &  &2.95 &\textbf{1.67} & 3.61 & 0\\
		  & 100 &  &  &1.22 &\textbf{1.06} & 12.0 & 0\\
		  & 125 &  &   &2.74 &\textbf{1.92} & 10.4 & 0\\
		  & 150 & &   &2.22 &\textbf{1.70} & 47.7 & 1\\
		  & 175 &   &  &\textbf{2.40} &7.45 & 127 & 3\\
		  & 200 &  &   &\textbf{3.04} &3.58&  162& 4\\	
		\hline
		30 & 30 & 0.25 & 45 &1.50 &0.525 & \textbf{0.240} & 0 \\ 		
		 & 60 &   &   &1.64 &\textbf{0.920} & 2.21 & 0 \\  		
		  & 90 &  &   &\textbf{1.01} &1.14 & 4.26 & 0 \\ 		
		  & 120 &   &   &1.28 &\textbf{1.03} & 15.8 & 0 \\
		  & 150 &   &   &0.910 &\textbf{0.730} & 6.30 & 0 \\ 
		  & 180 &   &    &2.92 &\textbf{1.98} & 74.0 & 0 \\ 				
		\hline
	\end{tabular}
\end{table}

We empirically observed that when the ratio of the number of users to the number of holes is large, our method works much more effective than its counterpart \their{}. 
Table~\ref{tab-ratio}, in which a total of 600 instances of the problem are solved (60 instances per row), also provides 
a strong evidence for this fact.
In this table,
the parameter
$ \delta $
is set as 
in \cite[Tables~1 and 2]{ilp}. 
In the first five
rows, we have $ M\approx 30 $,
$ N\in\{30,60,\ldots,150\} $,
and 
$ \delta_j\in[2R_j,3R_j] $ 
($ 1\leq j \leq N $).
In the next
five rows,
we have
$ M\approx 50 $,
 $ N\in\{50,100,\ldots,250\} $,
 and $ \delta_j\in[ 
 \max\{R_j, 15\}, \min\{2R_j, 40\}] $ ($ 1\leq j \leq N $).
For each row, 60 random instances are used.

\begin{table}[tbh]
	\caption{The effect of the ratio of the number of users to the number of holes on the time performance 
		(in seconds)
		of the two approaches. For each row, 60 random instances are used.
		The column labeled with ``TL'' reports the number of 
		CPLEX
		runs
		stopped due to the time limit
		(for solving \their{} models). None of the \sepi{} and \sepii{} runs reached the 600 seconds time limit.}
	\label{tab-ratio}
	\centering
	\begin{tabular}{|ll|c|cc|cc|}
		\hline
		$ M $& $ N $&$ \delta $ & \sepi{} & \sepii{} & \their{} & TL \\%& integral\\
		\hline
		$ \approx 30 $&$ 30 $&As in \cite[Tab.~1]{ilp} & 1.17 &2.67 &\textbf{0.293}  & 0 \\%&  36/60\\
		&$ 60 $ &  & 3.63 & \textbf{2.33} &   11.0 & 0  \\%&  31/60\\
		&$ 90 $&  & 3.98 & \textbf{1.49} &32.2  &  2 \\%&30/60 \\
		&$ 120 $&  & 21.3  &\textbf{7.38} & 87.8 & 6\\% & 35/60\\
		&$ 150 $&  &32.7  &\textbf{11.8} & 90.8 & 8 \\%& 30/60\\
		\hline
		$ \approx 50 $&$ 50 $ & As in \cite[Tab.~2]{ilp}& 0.338 & 0.577&  \textbf{0.150}  &  0  \\%&  34/60\\
		&$ 100 $&   &  0.773& \textbf{0.493} & 1.04 & 0 \\%& 38/60 &\\
		&$ 150 $ & &1.27 & \textbf{0.482} & 7.48 & 0 \\%& 38 / 60\\
		&$ 200 $&  & 0.802 & \textbf{0.715}& 27.8  & 1 \\%& 43 /60  \\
		&$ 250 $&  & \textbf{4.18} &5.27 &101 & 6 \\%&  35/60 \\
		\hline
	\end{tabular}
\end{table}

 The superiority of the proposed \bnp{} procedure
 stems from the strength
 (tightness) of the LP relaxation
 of the
 formulation (BILP).
 The LP relaxation
  of the formulation (BILP)
  can provide much tighter upper bounds
  than
  the LP relaxation of
  \their{}. 
 To see this fact more clearly,
 we reported the 
 optimal objective values of the
 LP relaxations of
 (BILP) and \their{} for the 20 instances
 corresponding to
 the 8th row of Table~\ref{tab1-h25},
 in which we have
 $ M=25 $, $ N=200 $, $ q=0.5 $,
 and $ \delta_j\in[2R_j,3R_j] $.
 Table \ref{tab-relax} reports these values.
 The numbers in parentheses are
 the
  absolute integrality gaps. As  can be seen, for a given instance,
   the LP relaxation of (BILP) 
   can
   provide a
    much sharper
    upper-bound than the LP relaxation
    of \their{}. 
    For example, for
    the 6th considered
    instance of the problem, the absolute 
    integrality gap
    for (BILP) is 0.05, and for \their{} is 24.435. 
These
sharp upper-bounds
allow for a much more
effective 
pruning of the search tree,
which is apparent in the
performance
of
\sepi{} and \sepii{}.

 \begin{table}[tbh]
	\centering
	\caption{%
	The 
	optimal objective values of the
	LP relaxations of
	(BILP) and \their{} for the 20 instances
	corresponding to
	the 8th row of Table~\ref{tab1-h25}.
	For these instances, we have
	$ M=25 $, $ N=200 $, $ q=0.5 $,
	and $ \delta_j\in[2R_j,3R_j] $.
	The numbers in parentheses are
	the
	absolute integrality gaps.}
	\label{tab-relax}
	\footnotesize
	\begin{tabular}{|llll|llll|}
		\hline
		& (BILP) & \their{} & Opt. &  & (BILP) & \their{} & Opt. \\
		\hline
		1& 190.917 (0.117)& 196.46 (5.66)& 190.8 &
		2&164.64 (0.04) &165.04 (0.44)&164.6 \\
		3& 182.7 (0)& 194.77 (12.07)& 182.7 &
		4&194.68 (0.08) & 195.16 (0.56)& 194.6\\
		5&196.15 (0.05) &197.41 (1.31)&196.1  &
		6&159.45 (0.05) &183.835 (24.435)& 159.4\\
		7&195.456 (0.056) &195.71 (0.31) &195.4  &
		8&191.467 (0.067) &196.83 (5.43)& 191.4\\
		9&184.567 (0.067) &194.78 (10.28)& 184.5 &
		10& 189.1 (0.1)& 194.78 (5.78)& 189\\
		11&170.3 (0) & 180.083 (9.783)& 170.3 &
		12&181.233 (0.33) & 195.88 (14.68)& 181.2\\
		13&197.05 (0.05) & 197.4 (0.4)& 197 &
		14&182.3 (0) & 190.45 (8.15)& 182.3\\
		15&171.75 (0.05) & 181.77 (10.07)& 171.7 &
		16&189.333 (0.033) & 194.74 (5.44)& 189.3\\
		17&184.083 (0.083) & 197.37 (13.37)&184 &
		18&193.85 (0.05) & 195.06 (1.26) & 193.8\\
		19&174.85 (0.05) &187.414 (12.614)& 174.8 &
		20&197.433 (0.033) & 197.98 (0.58)& 197.4\\
		\hline
	\end{tabular}
\end{table}

The parameter $ \delta $ has a major impact on the behavior  of \their{}.
As has been stated in
\cite{ilp},
as the value
of $ \delta $ grows, more variables and constraints
need to be employed in
\their{}, which
 increases the 
solution time.
We observed that,
as the value of $ \delta $ increases, the execution time of our method increases as well. 
However, when the value of $ \delta $ increases, our method shows 
 a
 much better performance than its counterpart.
 In Table~\ref{tab-delta}, we examine the effect of the value of
 $ \delta $
 on the performance of 
\sepi{},
\sepii{}, and \their{}. In this table, $ \delta $ takes
   the values 
30, 35, 40, and 45. 
In the first four
rows, we have $ M\approx 30 $ and
$ N=120 $, and in the next
four rows
$ M\approx 30 $ and $ N=150 $.
For each row, 60 random instances are used.
    As can be observed from the table, as the value of $ \delta $ increases,
    the
     performance difference becomes more
      pronounced. For example, for
      $ M\approx 30 $,
      $ N = 120 $,
      and 
       $ \delta = 45 $, the average running time for \sepi{} is 30, for
       \sepii{} 
        is 7.6, and for \their{} is 302.
        Moreover,
for 25 of the 60 runs corresponding to this row, CPLEX reached the time limit of 600 seconds when solving
\their{}, 
and returned a feasible,  and not necessarily optimal, solution. 

   \begin{table}[tbh]
	\caption{The effect of the value of $ \delta $ on the time performance 
		(in seconds)
		of the two approaches. For each row, 60 random instances are used.
		The column labeled with ``TL'' reports the number of 
		CPLEX
		runs
		stopped due to the time limit
		(for solving \their{} models). None of the \sepi{} and \sepii{} runs reached the 600 seconds time limit.}
	\label{tab-delta}
	\centering
	\begin{tabular}{|l|c|c|cc|cc|}
		\hline
		$M $ & $ N $ &  $ \delta $  & \sepi{} & \sepii{} & \their{} & TL\\% & integral\\
		\hline		
		$ \approx 30 $ & 120  & 30 &1.08&\textbf{0.380}  & 1.62 & 0\\% & 45 / 60\\
		&  &  35 &  \textbf{2.43} &2.60 & 27.2  &  2  \\%& 33 /60\\
		&  &  40 & 9.77 & \textbf{3.13} & 110 & 7 \\%& 20/60\\
		&  &  45 & 30 & \textbf{7.60} & 302  & 25 \\%& 18/60 \\
		\hline
		$ \approx 30 $ & 150  & 30 & 1.64 & \textbf{0.972} & 1.34 & 0 \\%& 44 / 60\\
		&  &  35 & 6.03 & \textbf{1.78}& 20.2  & 0 \\%& 39/60\\
		&  &  40 & 15.6 &\textbf{9.03} & 111 & 9 \\%& 25 / 60\\
		&   & 45 &21.5 &\textbf{8.23} &210 & 18  \\%& 22/60 \\		
		\hline
	\end{tabular}
\end{table}

The integer-feasible 
 but not
 necessarily
 optimal solutions
 that can be obtained
 during the \bnb{} process
 are also valuable.
When solving an instance of
 the ILP problem,
 sometimes we are
satisfied with
 a solution that
  is integer-feasible,
  but not necessarily optimal. 
 In these cases, we 
 can set a time-limit for the solver. We are satisfied with the best solution that the solver has encountered during this
 period.
  This 
  practice 
  is more common for
   large instances for which obtaining the optimal solution may be very time consuming.
 Table~\ref{tab-feas} shows
 the \sepi{}
 and \sepii{} can 
obtain better
feasible solutions
than \their{}.
In this table, only 90 seconds of time is given to the routines, and the best solution obtained is reported.
 In this table, we have considered 20 instances of the problem. 
 For these instances, we have
 $ M=70 $, $ N=70 $, $ q=0.25 $,
 and $ \delta_j\in[2R_j,3R_j] $.
  As can be seen, the solutions
  obtained by
  \sepi{} and \sepii{}
are of better quality than the
solutions of \their{}.

 We conclude this section by a brief comparison between the performance of 
 \sepi{} and \sepii{}.
 As can be observed from the
 Tables \ref{tab1-h25}--\ref{tab-feas},
 both of these \bnp{} procedures
 can provide a much better performance
 than \their{}.
 Generally,
 \sepii{} performs better
 than \sepi{}. 
 As can be seen from Table~\ref{tab-feas},
  the performance of the two routines in finding a good feasible solution as soon as possible was almost the same. However, we observed that, for some of the problem instances, \sepii{} performed much better at proving the optimality. In fact, for \sepii{}, \textit{lowering} the upper bound---%
  so that it can be
  confirmed that the current 
  incumbent is in fact optimal---%
  was much better during the \bnp{} process,
  for some of the 
  considered
  problem instances.
  This can be regarded as the reason for its better performance.

  \begin{table}[tbh]
 	\centering
 	\caption{
 		A comparison of the capability of the approaches to find 
 		feasible,
 		and not necessarily
 		optimal, solutions
 		in 90 seconds.
 		20 instances of the problem
 		have been considered.
 		For these instances, we have
 		$ M=70 $, $ N=70 $, $ q=0.25 $,
 		and $ \delta_j\in[2R_j,3R_j] $.}
 	\label{tab-feas}
 	\footnotesize
 	\begin{tabular}{|lllll|lllll|}
 		\hline
 		&\sepi{} & \sepii{} & \their{} & Opt. &
 		&\sepi{} & \sepii{} & \their{} & Opt. \\
 		\hline
 		1&\textbf{71.5}&\textbf{71.5} & \textbf{71.5}  &71.5   &
 		2&\textbf{73.6}&\textbf{73.6} & 70.2  &  73.6 \\
 		3&\textbf{72.1}&\textbf{72.1} & 69.5  &  72.1 &
 		4&71.6&\textbf{71.7} & 70.6  &  71.7 \\
 		5&\textbf{69.7}&\textbf{69.7} & \textbf{69.7}  & 69.7  &
 		6&\textbf{69.9}&\textbf{69.9} &67.9   & 69.9  \\
 		7&\textbf{75.4}&\textbf{75.4} &  \textbf{75.4} & 75.4  &
 		8&\textbf{72.5}&69.2 & 67.1   &  72.6 \\
 		9&\textbf{74.4}&72.4 & \textbf{74.4}  &  74.4 &
 		10&\textbf{62.8}&\textbf{62.8} &  62.6 & 62.8  \\
 		11&\textbf{71.8}&71.7 & 70.4  &  71.9 &
 		12&\textbf{70.1}&\textbf{70.1} &  67.7 &  70.1 \\
 		%70.7&019,70.7,109.053,10 & 67.5  &  70.7 \\
 		13&\textbf{75.4}&\textbf{75.4} & 74.9  &  75.4 &
 		%64.8&023,61.1,105.599,4 & 61.3  & 64.8  \\
 		14&\textbf{75.5}&\textbf{75.5} &  74.5 & 75.5  \\
 		15&70.5&\textbf{70.7} & 69.7  &  70.7 &
 		16&\textbf{67.2}&61.9 & 65.5  & 67.2  \\
 		17&\textbf{71.9}&71.8 & 71  & 72.9  &
 		18&74.7&\textbf{74.8} &  66.3 &  74.8 \\
 		19&\textbf{72.4}&70.3 & 72.1  & 72.4  &
 		20&\textbf{72.3}&\textbf{72.3} &  70.4 &  72.3 \\
 		\hline
 	\end{tabular}
 \end{table}
 
 \section{Conclusions}
 In this paper,
  we revisited the problem of 
  spectrum hole assignment to cognitive secondary
 users with different hardware limitations.
  A novel 0-1 ILP formulation has been proposed, which
 provides very tight LP relaxation bounds. Moreover, we devised two different \bnp{} 
 procedures
 to overcome the problem of 
 huge number of decision variables in
 the proposed formulation. Exhaustive numerical experiments demonstrate that the
 proposed approach can
 achieve a much better performance
  than the  best currently 
  available ILP formulation of the problem. Specifically, we reported
 some scenarios in which, the proposed
 approach was able to solve the problem instances 
 in about 0.02 of the
 time required by its counterpart.
 
 The work presented in this paper may 
 be extended in different directions. Other user-specific parameters, rather than the MAR and the required bandwidth, may be
 taken into account in the problem formulation. For instance, the users may have different preferences about the spectrum holes, based on
 how much data-rate they can gain from each hole (due to their locations and/or technologies).
 Another direction is to take into consideration other metrics (objective functions) such as \textit{fairness}, instead of considering only the spectrum utilization. Finally, incorporating learning mechanisms
 to predict the behavior of the PUs
 is a promising direction for future works.
 Based on analyzing the behavior of PUs,
  especially in scenarios where they may frequently
 change their spectrum usage pattern, we can
 seek for a more \textit{stable} assignment.
  We may first rank the holes based on 
  their
   availability probabilities, and then
   provide an acceptable
   hole assignment from the view point of stability.

 \section{Acknowledgements}
 This work was supported by by Iran's National Elites Foundation (INEF).

\bibliographystyle{elsarticle-num}

\end{document}